\documentclass[aps,prx,reprint,groupedaddress,amsmath,amssymb,superscriptaddress]{revtex4-2}

\usepackage{graphicx}
\usepackage{dcolumn}
\usepackage{bm}
\usepackage{hyperref}
\usepackage{color}
\usepackage{gensymb}

\begin{document}

\preprint{APS/123-QED}

\title{Surface Waves on Self-Complementary Metasurfaces:\\ Intrinsic Hyperbolicity, Dual-Directional Canalization and \\TE-TM Polarization Degeneracy }

\author{Vladimir Lenets}
\altaffiliation{These authors contributed equally to this work}
\affiliation{Department of Physics and Engineering, ITMO University, St. Petersburg, Russia}
\author{Oleh Yermakov}
\altaffiliation{These authors contributed equally to this work}
\affiliation{Department of Physics and Engineering, ITMO University, St. Petersburg, Russia}
\affiliation{V. N. Karazin Kharkiv National University, Kharkiv, Ukraine}
\author{Andrey Sayanskiy}
\affiliation{Department of Physics and Engineering, ITMO University, St. Petersburg, Russia}
\author{Juan~Baena} 
\affiliation{Physics Department, Universidad Nacional de Colombia, Bogota, Colombia}
\author{Enrica Martini} 
\affiliation{Department of Information Engineering and Mathematics, University of Siena, Siena, Italy}
\author{Stanislav Glybovski}
\email{s.glybovski@metalab.ifmo.ru}
\affiliation{Department of Physics and Engineering, ITMO University, St. Petersburg, Russia}
\author{Stefano Maci} 
\affiliation{Department of Information Engineering and Mathematics, University of Siena, Siena, Italy}

\date{\today}

\begin{abstract}
Self-complementary metasurfaces have gained significant attention due to their unique frequency-independent transmission and reflection properties and the possibility of the polarization transformation of plane waves. In this paper, we focus on the near-field spectrum to investigate both theoretically and experimentally the properties of surface waves supported by anisotropic self-complementary metasurfaces. We show that as a consequence of the electromagnetic Babinet's duality, such structure is hyperbolic for any frequency. We reveal the polarization degree of freedom inherent to plane waves by demonstrating the broadband TE-TM polarization degeneracy of the surface waves along two principal directions. We demonstrate the possibility of switching the canalization direction of $90^{\circ}$ by a very small frequency shift paving a way to the extreme tunability and surface wave routing. The results obtained open a plethora of opportunities for practical applications, including flat polarization devices, optical data processing systems and antennas.  
\end{abstract}

\maketitle


\section{\label{sec:level1} Introduction}

Metasurfaces have gained significant attention due to their ability to control the phase, amplitude and polarization of the incident electromagnetic (EM) waves in transmission and reflection~\cite{holloway2012overview,kildishev2013planar,yu2014flat}. Metasurfaces denote a thin-layer periodic array of subwavelength scatterers tailored to achieve the necessary goals in EM waves control for a number of practical applications such as lenses, antennas, absorbers, filters, polarizers, holograms, etc~\cite{glybovski2016,chen2016review}. Even a more promising fact is that metasurfaces provide the ultrathin platform to implement on-chip and photonic devices based on near-field effects and in-plane propagation of surface plasmon-polaritons with a plethora of perspective applications in communications, sensing and optical networks~\cite{gomez2016flatland,droulias2019surface}. Another important application area is the microwave and millimeter-wave antennas, where surface waves propagating at spatially non-uniform impedance metasurfaces have been used. Antennas can be realized by converting slow surface waves into radiating waves by introducing a periodic or quasi-periodic spatial modulation of the surface impedance~\cite{mts_add}. This method was successfully employed to design low-profile and high-gain directive antennas with radiation pattern shapes precisely controlled by the impedance modulation \cite{SaT_hol,Spr_lwa,cp_isoflux}. Furthermore, Luneburg \cite{printed_luneburg,nonun_lunenburg,thin_luneburg} and Maxwell fish-eye \cite{fisheye_lunenburg} flat microwave lenses were based on engineered curvilinear propagation of surface waves over a spatially non-uniform metasurface.

Particular interest is addressed to hyperbolic plasmons -- surface waves localized at hyperbolic metasurfaces~\cite{yermakov2015hybrid,gomez2015hyperbolic,mencagli2015surface,nemilentsau2016anisotropic}. For the last decade many exciting regimes of hyperbolic plasmons propagation were discovered and analyzed including negative refraction~\cite{high2015visible}, plasmon steering~\cite{stein2010surface,sinev2017chirality}, zero-index~\cite{liberal2017near}, unidirectional~\cite{gangaraj2019unidirectional,nemilentsau2019switchable} and canalization~\cite{correas2017plasmon} regimes of propagation. Despite the great success in the hyperbolic plasmons research, the potential of their further development and application is strictly limited by the narrow operational frequency range. 

Canalization \footnote{It is also known as \textit{collimation}, \textit{guiding}, \textit{chanelling}, \textit{tunneling}} regime is a well-known phenomenon in photonic crystals~\cite{kosaka1999self,witzens2002self} and metamaterials~\cite{belov2005canalization,shen2016metamaterial}. It leads to the enhanced resolution (greater than diffraction limit), imaging and lensing~\cite{ikonen2006experimental,belov2006subwavelength,silveirinha2007subwavelength}. Recently, the study of surface waves canalization regime at anisotropic metasurface has gained a great importance due to the enormous application potential in flat devices, on-chip networks and optical signal control~\cite{stein2012self,gomez2016flatland,correas2017plasmon}. This phenomenon is observed in the extremely anisotropic systems in the vicinity of the near-zero regimes. This exciting effect is still poorly used and remains exotic in the aspect of applications due to the single frequency and unidirectional operational regime. 

 \begin{figure}[htbp]
  \centering
  \includegraphics[width=0.9\linewidth]{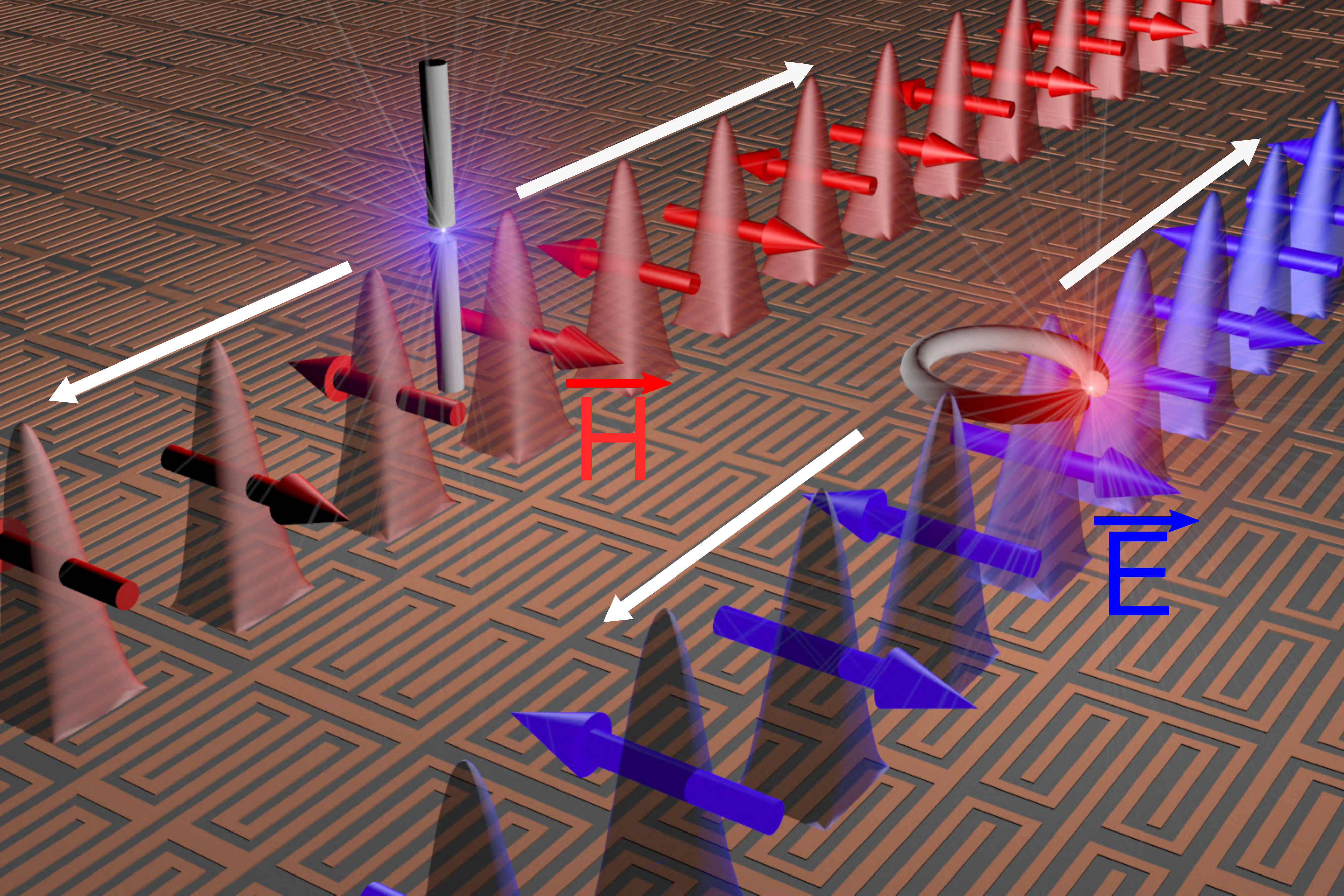}
  \caption{State-of-art picture of the canalized and polarization-degenerate surface waves at the self-complementary metasurface under study. The \textcolor{red}{red} and \textcolor{blue}{blue} arrows demonstrate the instantaneous direction of \textcolor{red}{magnetic} and \textcolor{blue}{electric} fields of \textcolor{red}{TM} and \textcolor{blue}{TE} surface plasmons excited by vertical \textcolor{red}{electric (probe)} and \textcolor{blue}{magnetic (loop)} dipole-like sources, respectively. The conical shapes schematically show the field amplitude sharply decreasing with distance from a metasurface. The white arrows correspond to the wave propagation directions emulating the canalization propagation regime of the surface plasmon-polaritons. }
  \label{fig0}
\end{figure}

The surface waves at anisotropic metasurfaces possess hybrid TE-TM polarization~\cite{yermakov2016spin,yermakov2018experimental} providing an efficient tool for the spin-dependent electromagnetic phenomena~\cite{hayat2015lateral,yang2017hyperbolic}. However, despite the complicated polarization structure of surface waves, the development of flat optics is significantly limited by the inefficient polarization control of the propagating localized EM waves. Except for accidental intersections at several points, the dispersion curves of TE and TM surface modes are not degenerate in sharp contrast to the plane waves in isotropic medium. Nevertheless, polarization degeneracy of the TE and TM modes is the fundamental operational principle of the classical bulk polarizers. The polarization transformation of propagating surface waves also requires the broadband TE-TM polarization degeneracy of the eigenmodes spectrum inherent to the bulk waves.

In this work, we show that canalization and broadband polarization degeneracy can be achieved with resonant and anisotropic self-complementary metasurfaces. In particular, it is shown that both TE and TM surface waves have the same dispersion and can be canalized as schematically depicted in Fig.~\ref{fig0}. 

Anisotropic self-complementary metasurfaces are single-layer metal patterns that remain the same after Babinet inversion except for some translation or rotation. They were shown to operate as frequency-selective filters~\cite{ortiz2013self}, perfect absorbers~\cite{urade2016broadband}, LP-to-CP converters for incident plane waves \cite{SC_LP_CP, baena2017broadband} and they have frequency-constant transmission properties when excited by circularly-polarized waves \cite{freq_ctrl_pol}. In contrast to isotropic checkerboard which has self-complementary patterns requiring perfectly sharp corners of planar square conductors \cite{compton1984babinet,nakata2013plane,chck_scr}, their anisotropic counterparts are constructed of alternating inductive and capacitive complementary strips, and therefore, are easily to implement. In~\cite{gonzalez2015surface} it was shown that two mutually-complementary surfaces, one with inductive and the other with capacitive impedance, have the same dispersion of TM and TE surface waves, respectively. However, surface waves on anisotropic self-complementary metasurfaces have not been previously studied. Here, based on the intrinsic properties of resonant self-complementary metasurfaces we demonstrate both theoretically and experimentally (i) the existence of hyperbolic plasmons within an extremely wide frequency range, (ii) the possibility to switch the direction of canalization by a very small frequency shift, and (iii) the broadband TE-TM polarization degeneracy of surface waves.

\section{Formulation}

\subsection{Problem statement}

\begin{figure}[b]
  \centering
  \includegraphics[width=0.9\linewidth]{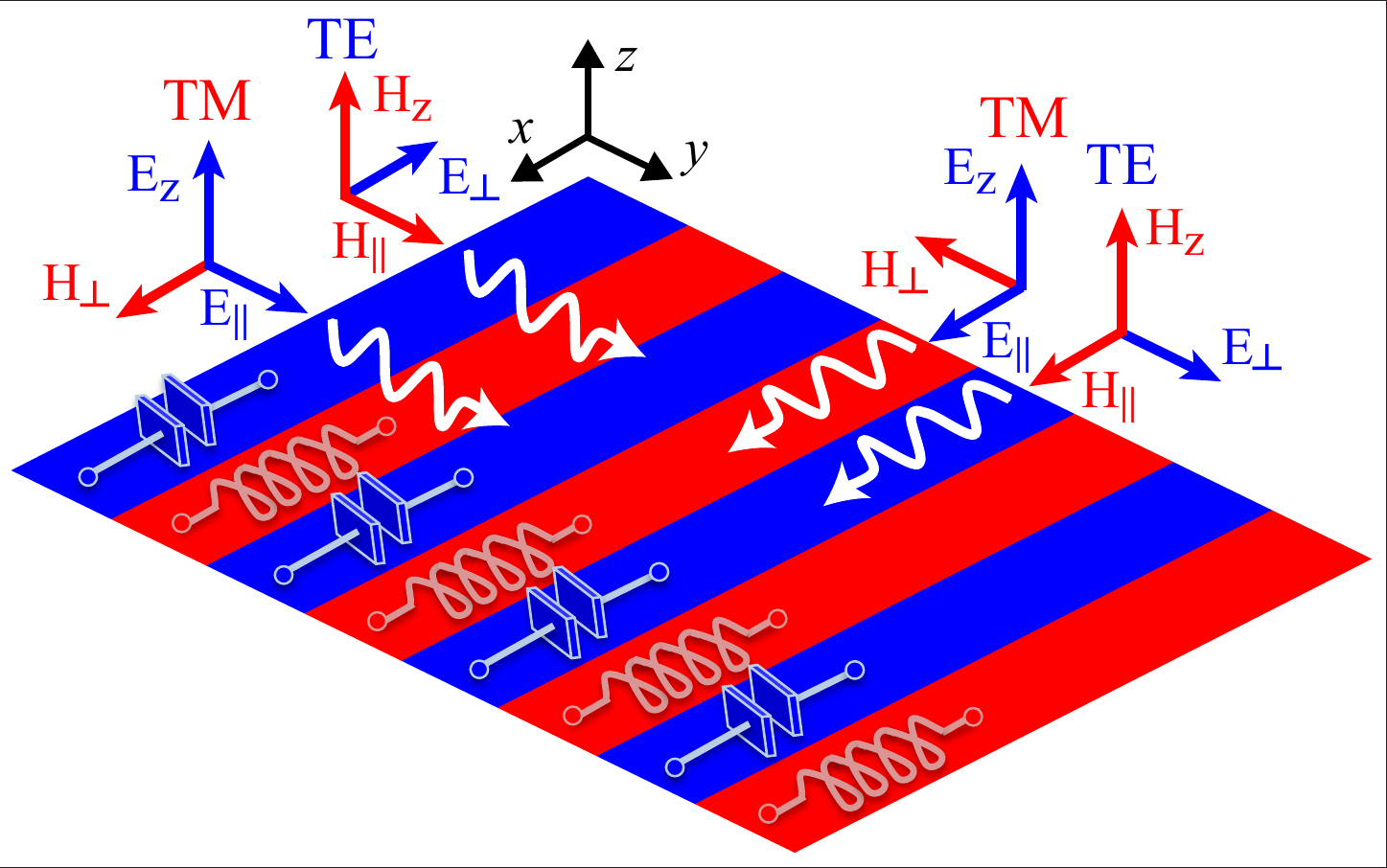}
  \caption{Geometry of the problem constituted by a collections of  co-planar inductive and capacitive parallel strips of a width small in terms of a wavelength immersed in free space. The polarization of TE and TM surface wave modes with propagation direction both parallel and orthogonal to the strips is also depicted.  }
  \label{fig1}
\end{figure}

We consider a single-layer resonant anisotropic metasurface floating in free space constituted by alternance of infinitely thin complementary inductive and capacitive strips with narrow width in terms of the operating wavelength (Fig.~\ref{fig1}), so that they can be homogenized with boundary condition
\begin{equation}
    \mathbf{J} = \hat{\mathbf{n}} \times \left( \mathbf{H_1 - H_2} \right) = \hat{Y} \mathbf{E}_{\tau}.
    \label{BC1}
\end{equation}
In~\eqref{BC1} $\hat{\mathbf{n}}$ is the unit normal vector from the lower to the upper half-space, $\mathbf{J}$ is a surface electric current density proportional to the jump of the tangential component of the magnetic field across the metasurface plane, $\mathbf{E}_{\tau}$ is a tangential electric field component continuous across the metasurface, $\hat{Y}$ is the local effective surface admittance tensor describing the anisotropic metasurface. 

It is worth noting that the above structure respects the Babinet's duality principle if the strips are mutually complementary. This can be obtained by realizing geometrical complementary slots and dipoles along the strips still maintaining them compact in terms of a wavelength. If two half-spaces above and below are filled with different permittivities, the structure, strictly speaking, does not respect the duality. Nevertheless, the physical mechanism we are going to describe will be the same. First, we analyze an ideally dual case, while in the experiment we implement a metasurface as a self-complementary copper pattern on a thin dielectric substrate. As it will be shown, in the considered practical metasurface realization, the Babinet's duality principle is fulfilled still well.


\subsection{Admittance tensor}

The anisotropic metasurfaces as well as other ultrathin subwavelength systems (graphene, van der Waals materials, 2D electron gas, etc.) can be reasonably described in terms of the effective surface admittance (conductivity) approach. We assume a metasurface creates no cross-polarization when the incident plane wave has only $x$- or $y$-directed tangential electric field. This means, in these coordinates the tensor can be diagonalized along the principal axes as follows
\begin{equation}
    \hat{Y}_0 = \begin{pmatrix}
    Y_{x} & 0 \\
    0 & Y_{y}
    \end{pmatrix}.
    \label{Y0tensor}
\end{equation}

The resonant response of a metasurface can be expressed by a Lorentzian form of one of the admittance tensor components (for example, here we have chosen $Y_y$ component):
\begin{equation}
  Y_{y} = \sum_j \frac{i Y_0 N_{j} \omega^2}{\omega^2 - \Omega_{j}^2 + i \gamma_{j} \omega}.
  \label{Lorentz}
\end{equation}
Here, $j$ denotes the resonance number, $Y_0 = 1/Z_0$, $Z_0$ is the free space wave impedance, $N_j$ is a non-dimensional normalization factor, $\Omega_j$ is the resonance angular frequency, $\gamma_j$ is the resonance bandwidth.

\begin{figure*}[htbp]
  \centering
  \includegraphics[width=0.98\linewidth]{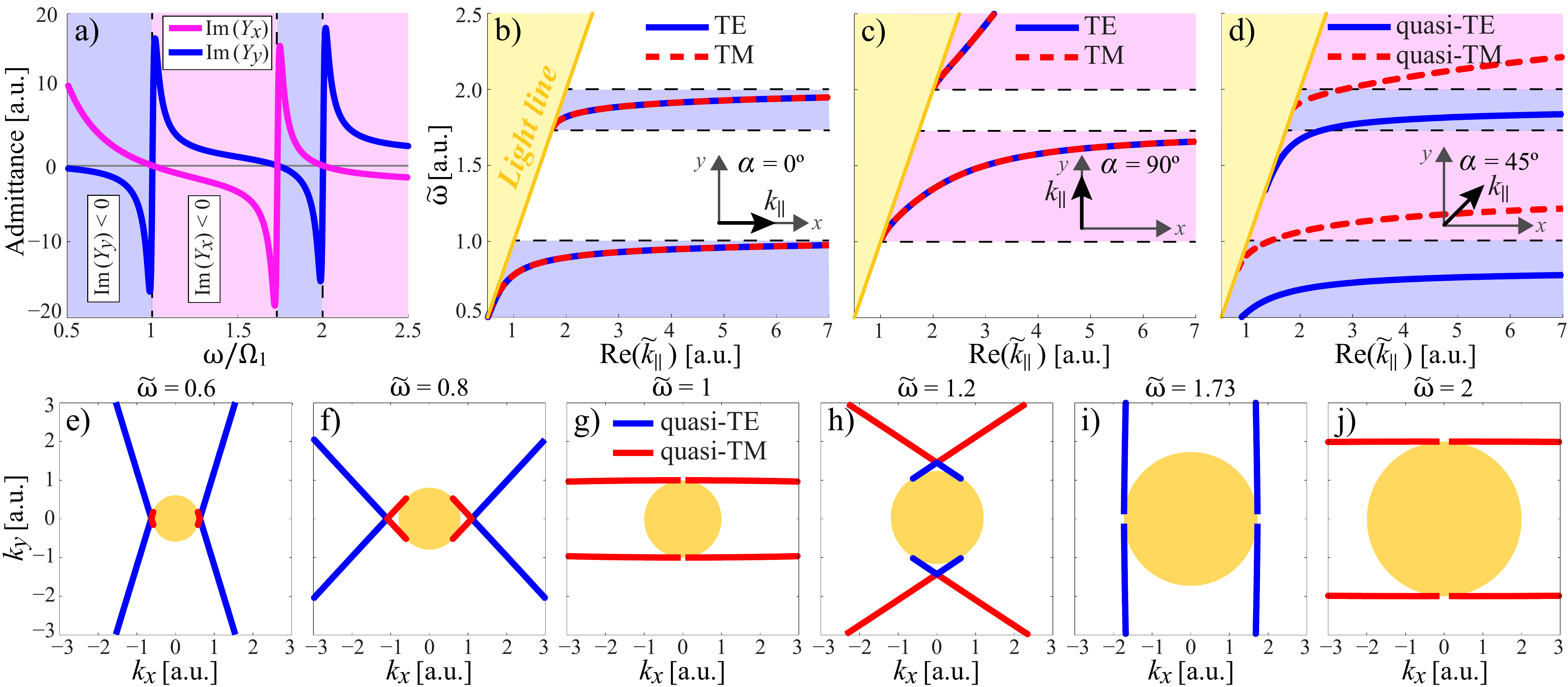}
  \caption{Analytically calculated properties of an ideal resonant self-complementary metasurface: (a) Frequency dependence of the imaginary parts of admittance tensor components. Lorentzian parameters are $N_{1} = 1$, $\widetilde{\Omega}_1 = 1$, $N_{2} = 0.5$, $\widetilde{\Omega}_2 = 2$, $\widetilde{\gamma}_1 = \widetilde{\gamma}_2 = 0.03$.  (b-d) Dispersion of surface waves localized at self-complementary metasurface in different directions $\alpha = 0^\circ$ (b), $90^\circ$ (c), $45^\circ$ (d). (e-j) Isofrequency contours of quasi-TE and quasi-TM surface modes in $(k_x = \widetilde{k}_{\|} \cos{\alpha},\, k_y = \widetilde{k}_{\|} \sin{\alpha})$ space at the angular frequencies $\widetilde{\omega} = 0.6$ (e), $\widetilde{\omega} = 0.8$ (f), $\widetilde{\omega} = 1$ (g), $\widetilde{\omega} = 1.2$ (h), $\widetilde{\omega} = 1.73$ (i), $\widetilde{\omega} = 2$ (j). Here, we use the dimensionless units $\widetilde{\Omega} = \Omega / \Omega_1$, $\widetilde{\omega} = \omega / \Omega_1$, $\widetilde{\gamma} = \gamma / \Omega_1$, $\widetilde{k}_{\|} = c k_0 / \Omega_1 = \widetilde{\omega}$.  }
  \label{fig2}
\end{figure*}

According to the Babinet's duality relation the surface admittance tensor components obey the relation~\cite{baena2017broadband}:
\begin{equation}
  \text{det}\hat{Y}_0 = Y_{x} Y_{y} = 4 Y_0^2,
  \label{complementary_cond_Y}
\end{equation}
So, the opposite admittance tensor component $Y_x$ can be strictly determined by the relation~\eqref{complementary_cond_Y} and is inversely proportional to the one defined in Eq.~\eqref{Lorentz}. We note that the duality relation~\eqref{complementary_cond_Y} highlights a cancellation between the poles and the zeros of the admittances (Fig.~\ref{fig2}a), that implies a frequency alternance between series and parallel type resonances for the two eigenvalues of the tensor. 



In order to find the expression of the admittance in an arbitrary reference system, it is useful to introduce a wavevector-fixed coordinate system constituted by unit vectors parallel and orthogonal to the direction of propagation. Assuming that the wave vector forms an angle $\alpha$ with respect to the $x$-axis, in this reference system, the admittance tensor may be written as   
\begin{equation}
    \hat{Y} = \begin{pmatrix}
    Y_{\|, \|} & Y_{\|, \bot} \\
    Y_{\bot, \|} & Y_{\bot, \bot}
    \end{pmatrix},
\end{equation}
where
\begin{equation}
\begin{split}
    & Y_{\|, \|} = Y_{x} \cos^2{\alpha} + Y_{y} \sin^2{\alpha} , \\
    & Y_{\bot, \bot} = Y_{x} \sin^2{\alpha} + Y_{y} \cos^2{\alpha} , \\
    & Y_{\|, \bot} = Y_{\bot, \|} = (Y_{x} - Y_{y}) \cos{\alpha} \sin{\alpha}.
    \label{Yreference}
\end{split}
\end{equation}
When $\alpha$ is zero, \eqref{Yreference} recovers \eqref{Y0tensor}, when it is $90^\circ$, the tensor becomes again diagonal with swapped position of the diagonal eigenvalues with respect to $\alpha =0$. In this system (see Fig.~\ref{fig1}), the resonances change their roles with respect to propagation along the strips, namely the parallel resonances become series resonances and vice versa, as mentioned before.  


\subsection{Dispersion equation}

The dispersion equation of surface waves at an anisotropic metasurface described with the effective local surface admittance tensor can be found analytically~\cite{yermakov2015hybrid}:
\begin{equation}
\begin{split}
    \left( \frac{\varepsilon_1 k_0}{\kappa_1} + \frac{\varepsilon_2 k_0}{\kappa_2} + i \widetilde{Y}_{\|, \|} \right)  \left( \frac{\kappa_1}{k_0} + \frac{\kappa_2}{k_0} - i \widetilde{Y}_{\bot, \bot}  \right)  = \widetilde{Y}_{\|, \bot} \widetilde{Y}_{\bot, \|}.
    \label{disp_eq}
\end{split}
\end{equation}
Here, $k_0 = \omega/c$, $\kappa_{1,2} = \sqrt{k_{\|}^2 - \varepsilon_{1,2} k_0^2}$ is a penetration depth, $k_{\|}$ is the surface wave propagation constant, $\varepsilon_1$ and $\varepsilon_2$ are the dielectric constants for $z$ positive and negative, respectively. In~\eqref{disp_eq} and further, we use the following admittance normalization $\widetilde{Y} = Y / Y_0$. 

One can notice that first and second factors in the left side of Eq.~\eqref{disp_eq} correspond to the dispersion laws of TM and TE modes, respectively, while the right side of Eq.~\eqref{disp_eq} contains a coupling factor related to anisotropy. So, the spectrum of an anisotropic metasurface consists of the modes with hybrid TE-TM polarization, called usually quasi-TM and quasi-TE ones~\cite{yermakov2015hybrid,yermakov2018experimental} depending on which polarization component is dominant. Importantly, the surface modes possess purely orthogonal TE and TM polarizations in the main axes directions $\alpha = n \pi/2$, where $n$ is an integer. For any different propagation angle $\alpha \neq n \pi/2$ the spectrum represents the set of the hybrid TE-TM surface waves.

The dispersion equation~\eqref{disp_eq} can be solved analytically in the symmetric case $\varepsilon_1 = \varepsilon_2 = \varepsilon$:
\begin{equation}
    \kappa^{\text{TM,TE}} = k_0 \frac{\zeta \pm \sqrt{\zeta^2 - \varepsilon \widetilde{Y}_{\|, \|} \widetilde{Y}_{\bot, \bot} }}{-i \widetilde{Y}_{\|, \|}}, \; \zeta = \varepsilon + \frac{\text{det}\hat{\widetilde{Y}}_0}{4},
    \label{disp_eq_kappa_sol}
\end{equation}
where square root is defined with positive real part and the upper (lower) sign corresponds to TM (TE) mode. In absence of losses, the term under square root becomes real and positive with $\zeta = \varepsilon + 1$. In free space we can express the dispersion of the in-plane complex wavevector component of the surface waves localized at self-complementary metasurface by applying the Babinet's duality relation~\eqref{complementary_cond_Y} as follows:
\begin{equation}
    k_{\|} = k_0 \sqrt{  1 - \left(\frac{2 \pm \sqrt{-\widetilde{Y}_{\|, \bot} \widetilde{Y}_{\bot, \|}}}{\widetilde{Y}_{\|,\|}}  \right)^2 }.
    \label{disp_eq_sol}
\end{equation}
This equation is specialized for the propagation direction along the strips ($\alpha = 0^\circ$), orthogonal to the strips ($\alpha = 90^\circ$) and at $\alpha = 45^\circ$, respectively, thus leading to
\begin{equation}
\begin{split}
    &k_{\|}^{\alpha = 0^\circ} = \eta_x k_0 \sqrt{1 - \frac{\widetilde{Y}^2_y}{4} } = \eta_x k_0 \sqrt{1 - \frac{4}{\widetilde{Y}^2_x} }, \\
    &k_{\|}^{\alpha = 90^\circ} = \eta_y k_0 \sqrt{1 - \frac{\widetilde{Y}^2_x}{4} } = \eta_y k_0 \sqrt{1 - \frac{4}{\widetilde{Y}^2_y} },\\
    &k_{\|}^{\alpha = 45^\circ} = k_0 \sqrt{  1 - \left( \frac{4 \pm \sqrt{-\widetilde{Y}_{-}^2}}{\widetilde{Y}_+} \right)^2},
\end{split}
 \label{kx_anal}
\end{equation}
where $\eta_{x,y} = U[\text{Im}(Y_{x,y})]$ and $U[\; ]$ is the Heaviside unit step function, $\widetilde{Y}_{+,-} = \widetilde{Y}_x \pm \widetilde{Y}_y$, the upper (lower) sign for $\alpha = 45^\circ$ corresponds to quasi-TM (quasi-TE) modes. This relation states that the eigenmodes propagating along $x$- and $y-$axis exist only within the frequency range corresponding to $\text{Im}\left( Y_y \right) < 0$ and $\text{Im}\left( Y_x \right) < 0$, respectively. These propagation constants are plotted in Figs.~\ref{fig2}b-\ref{fig2}d. We note that in the two principal directions the degeneracy of the propagation constant for TE and TM waves takes place, while at $45^\circ$ the modes exhibit the different wavenumbers.

It is important to point out that, strictly speaking, any substrate with permittivity different from the permittivity of free space violates the Babinet's duality relation~\eqref{complementary_cond_Y}. Nevertheless, the dispersion equation~\eqref{disp_eq} remains relevant for any environment and it could be solved analytically for small difference between superstrate and substrate permittivies using a perturbation theory~\cite{yermakov2015hybrid}. We will further demonstrate both numerically and experimentally the consistence of the analytically predicted results for the self-complementary metasurface at thin dielectric substrate.


\subsection{Isofrequency contours}

One of the powerful tools to analyze the wave propagation features is the isofrequency contours (IFCs). The possible isofrequency contours for the self-complementary metasurface described by the surface admittance tensor, shown in Fig.~\ref{fig2}a, are presented in Figs.~\ref{fig2}e-\ref{fig2}j. One can follow the evolution from the horizontal (Figs.~\ref{fig2}e,\ref{fig2}f) and vertical (Fig.~\ref{fig2}h) hyperbolic to the horizontal (Figs.~\ref{fig2}g,\ref{fig2}j) and vertical (Fig.~\ref{fig2}i) flat IFCs. The hyperbolic-like and emerging in the vicinity of the resonances flat IFCs correspond to the hyperbolic and canalization regime of a metasurface, respectively. The horizontal flat IFCs are observed in the vicinity of $Y_y$ resonances, while $Y_x$ is nearly zero, so the condition $|Y_y/Y_x| \gg 1$ is fulfilled. For vertical flat IFC the opposite situation takes place.

One can notice that for any frequencies except the resonances  the quasi-TE and quasi-TM IFCs have the intersection point along the corresponding main axes in accordance to Figs.~\ref{fig2}b,\ref{fig2}c. Its absence in the near-resonance frequency range corresponding to the flat IFCs is associated with the losses ($\gamma_1 = \gamma_2 = 0.03 \Omega_1$).

\section{Numerical analysis and measurements}

\subsection{Unit cell geometry and numerical analysis}

\begin{figure}[htbp]
  \centering
  \includegraphics[width=0.62\linewidth]{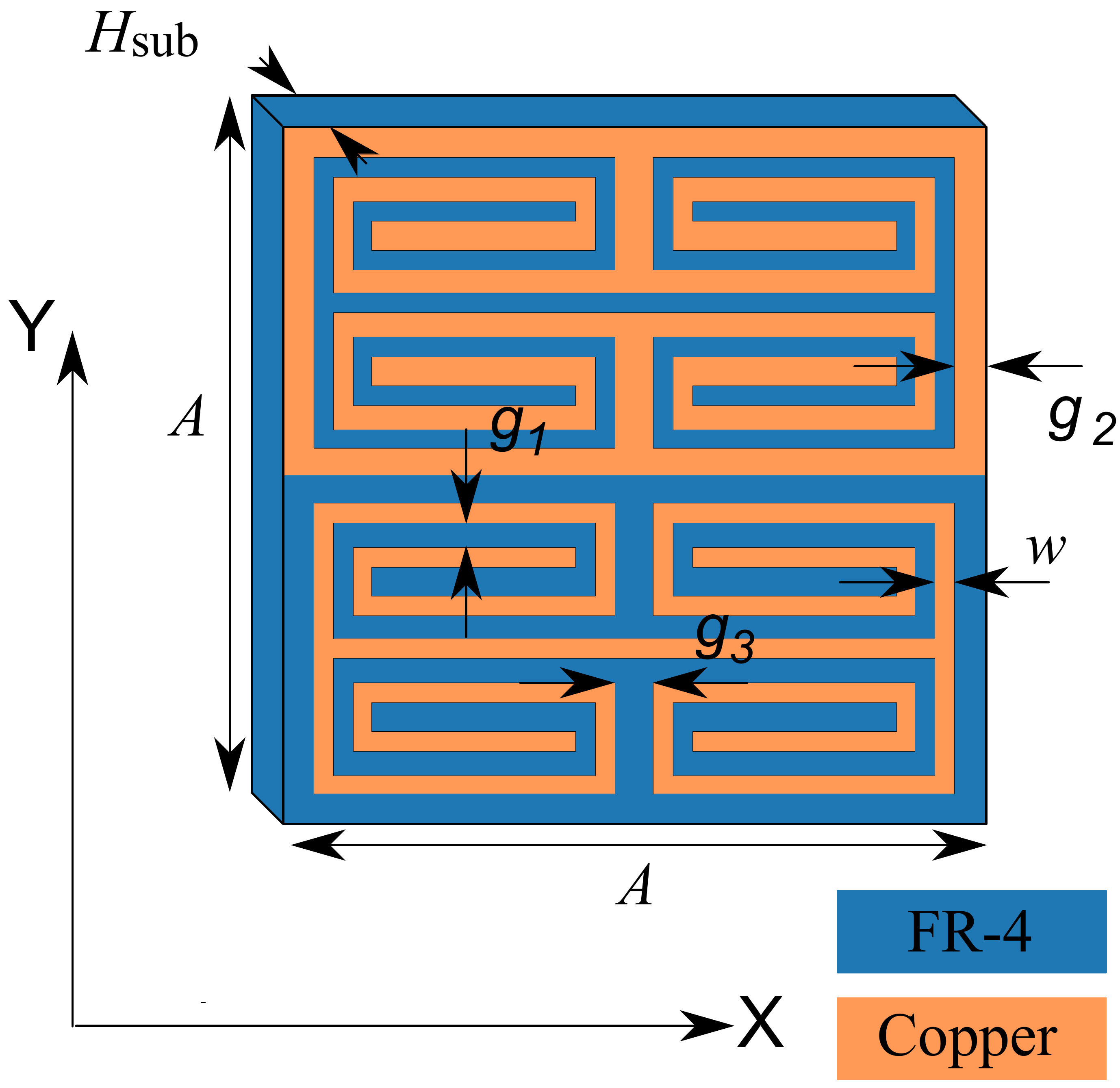}
  \caption{Unit-cell design of the resonant self-complementary metasurface under investigation.}
  \label{figUC}
\end{figure}

The unit-cell design of the considered metasurface is based on a meandered dipole and a complementary meandered slot (Fig.~\ref{figUC}). By numerical simulations the following geometric parameters were chosen to get the first resonance of the metasurface at around 5~GHz: periodicity of the square unit cell $A=7$~mm,  substrate thickness $H_\text{sub} =1$~mm, gap and metal strip widths $g_1 =0.25$~mm, $g_2 = 0.4$~mm, $g_3 = 0.2$~mm and $w =0.2$~mm.

\begin{figure}[htbp]
  \centering
  \includegraphics[width=0.99\linewidth]{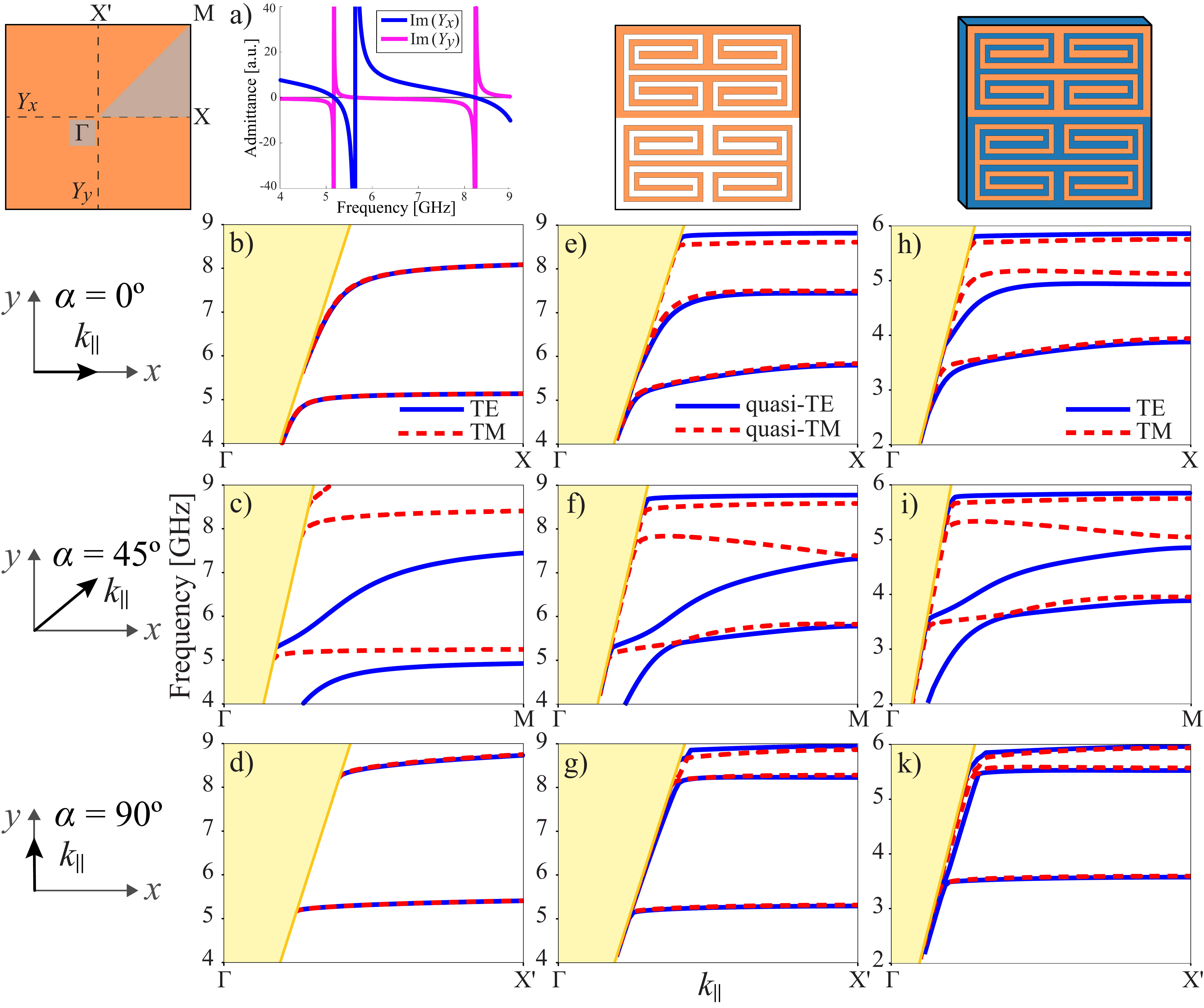}
  \caption{Dispersion diagrams for surface waves localized at the self-complementary metasurface, shown in Fig.~\ref{figUC}, without substrate (b-g) and with substrate (h-k) for different propagation angles (b,e,h) $\alpha = 0^\circ$, (c,f,i) $\alpha = 45^\circ$, (d,g,k) $\alpha = 90^\circ$. The dispersion curves were calculated analytically (b-d) by using the recovered effective surface admittance tensor (a) and with CST Eigenmode solver (e-k).
  }
  \label{fig4}
\end{figure}

The numerical simulation of dispersion curves and field patterns over the metasurface, shown in Fig.~\ref{figUC}, were done in Eigenmodes and Transient Solvers of CST Microwave Studio, respectively. The surface waves dispersion comparison between theory proposed in Section~II and numerical simulation is shown in Fig.~\ref{fig4}. Semi-analytical eigenmodes dispersion curves (Figs.~\ref{fig4}b-\ref{fig4}d) are based on the solution of Eq.~\eqref{disp_eq}, whereas the effective surface admittance tensor~\eqref{Y0tensor} of the real structure (Fig.~\ref{figUC}) without substrate was extracted from the numerically calculated S-matrix (Fig.~\ref{fig4}a)~\cite{pozar2009microwave}. On the other hand, we have found numerically the eigenmodes spectrum of the same structure without (Figs.~\ref{fig4}e-\ref{fig4}g) and with (Figs.~\ref{fig4}h-\ref{fig4}k) a dielectric substrate FR-4. The cross-check analysis proves the relevance of our formulation even in a case where the Babinet's duality is not rigorous, namely in the presence of the dielectric substrate. The surface waves spectra shown in Fig.~\ref{fig4} are in reasonable agreement both for the semi-analytical and numerical results. One can notice, however, that the characteristic frequencies in numerical simulation are slightly lower than the ones obtained by the semi-analytical process, which is probably consequence of the non-locality~\cite{gorlach2014effect,correas2015nonlocal,yermakov2018effective}.

\subsection{Measurements}

\begin{figure}[htbp]
  \centering
  \includegraphics[width=0.77\linewidth]{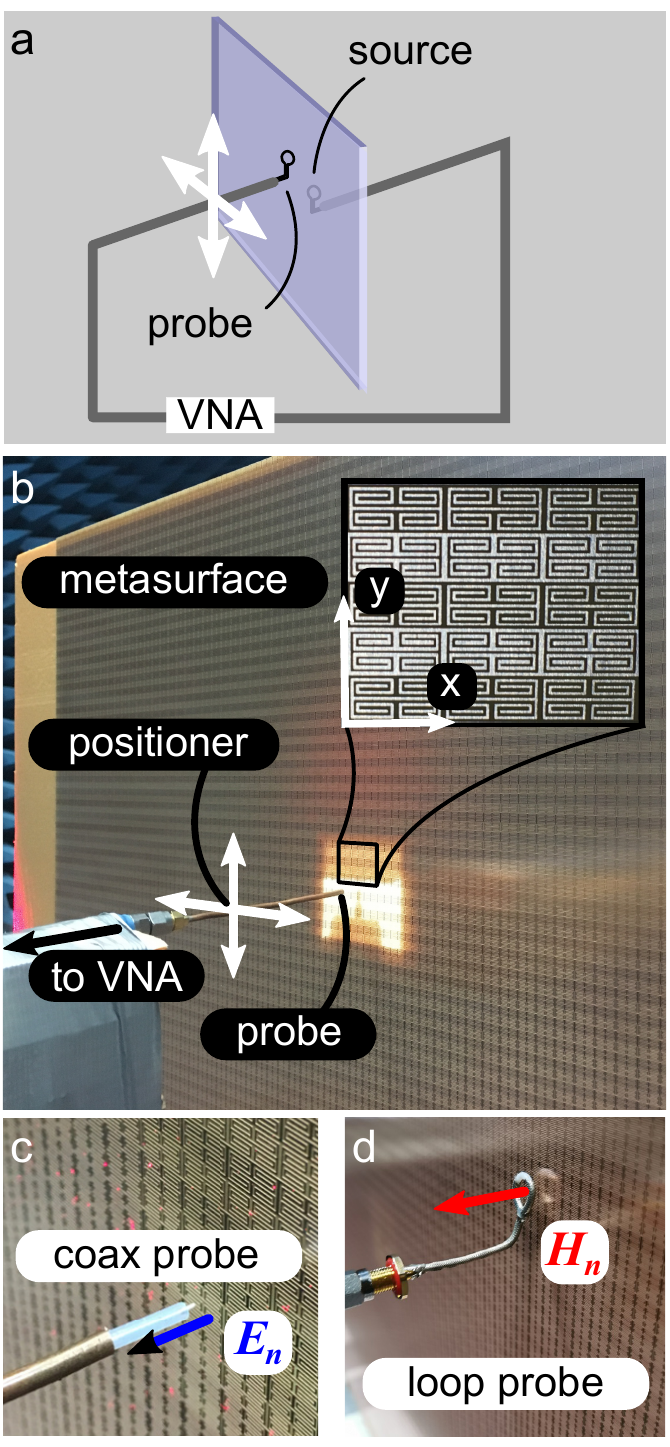}
  \caption{Scheme (a) and photo (b) of the field measurements setup. Electric (c) and magnetic (d) field probes.}
  \label{fig3}
\end{figure}

 The experimental sample was produced using a printed circuit board method on substrate FR-4 with the relative permittivity  $\varepsilon =3.9$ and loss tangent $\tan(\delta) = 0.02$. The full-size sample has dimensions $69 \times 55$ unit-cells or $483 \times 385$~mm$^2$. The manufactured sample was fixed using a foam substrate and was placed between a source antenna and a probe for near-field measurements in an anechoic chamber (Figs.~\ref{fig3}a,~\ref{fig3}b). The feeding probe during the measurements was fixed and placed in the middle of the structure at a distance of five millimeters from the sample. At the same time, a scanning probe was moved by a near-field scanner across the plane parallel to the metasurface on the opposite side of the structure such that the gap between the probe and sample was equal to ten millimeters. The probe was connected to a vector network analyzer (VNA) Agilent E8362C.

Two complex field maps have been obtained by two independent measurements for quasi-TE and quasi-TM surface waves, respectively. For the measurement of the quasi-TM surface wave, electric monopoles made of an open-ended coaxial cables (Fig.~\ref{fig3}c) were used both as a source and as a probe to measure the normal component of the electric field. For the quasi-TE surface waves measurement, Faraday loops \cite{jackson_classical_1999,pendry1999magnetism} were manufactured from coaxial cables (Fig.~\ref{fig3}d) and similarly used as transmitting and receiving probes to measure the normal component of the magnetic field. 
After the field maps were obtained, they have been converted to the isofrequency contours using space Fourier transformation, as it was proposed in \cite{Dockrey2016DirectOO,yang2017hyperbolic}. In our work, we have used fast Fourier transformation with zero padding and Hamming window to increase the resolution of the isofrequency contours.

\begin{figure*}[htbp]
  \centering
  \includegraphics[width=0.99\linewidth]{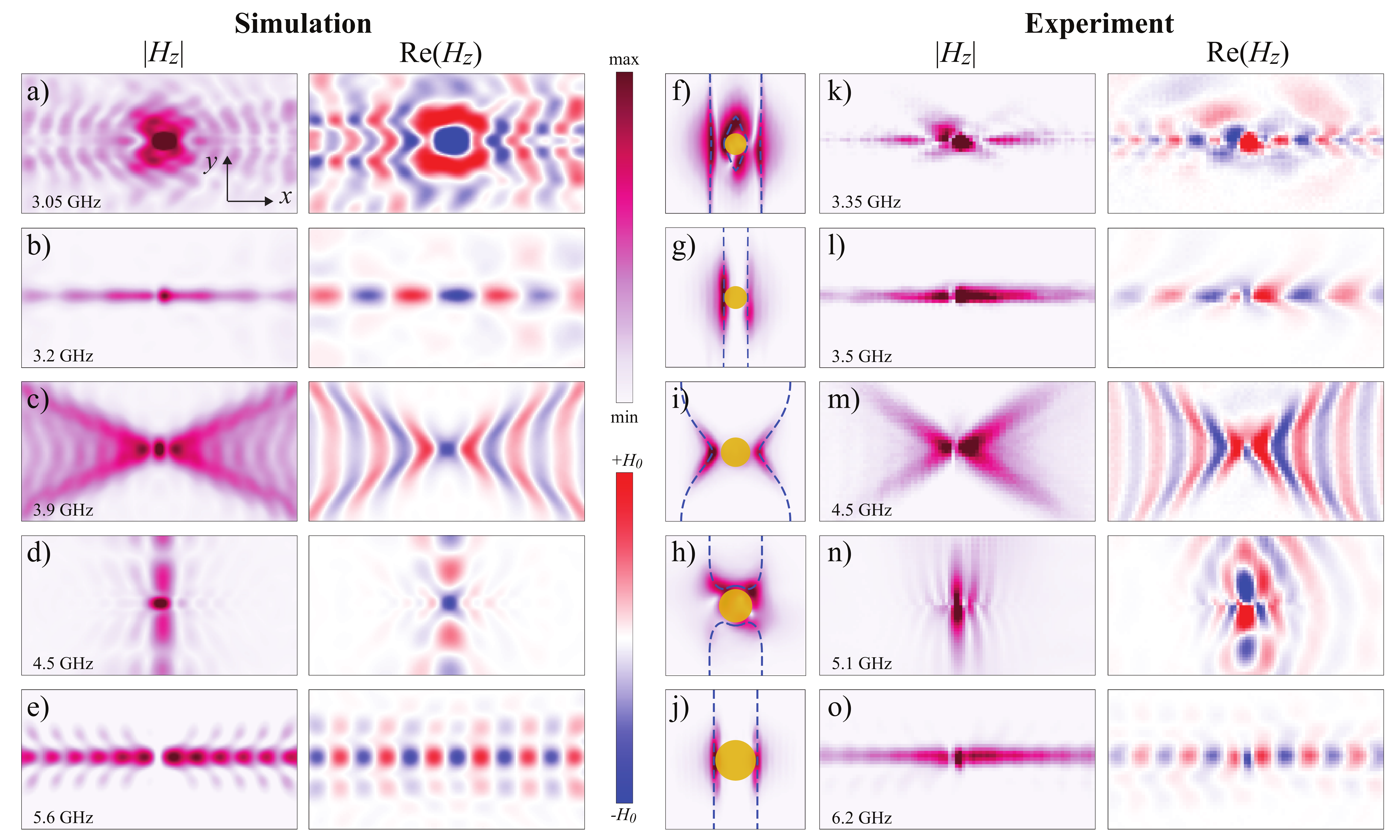}
  \caption{The spatial distribution of the absolute value (first and fourth columns) and real part (second and fifth columns) of the normal magnetic field component calculated numerically (a-e) and measured (k-o). The scanning area of the structure consists of 48$\times$24 unit cells (336$\times$168~mm$^2$). The central column (f-j) corresponds to the isofrequency contours within the first Brillouin zone restored from the measurements (color maps) and calculated numerically with CST Eigenmode solver (blue dashed lines). }
  \label{fig5}
\end{figure*}

\begin{figure}[htbp]
  \centering
  \includegraphics[width=0.99\linewidth]{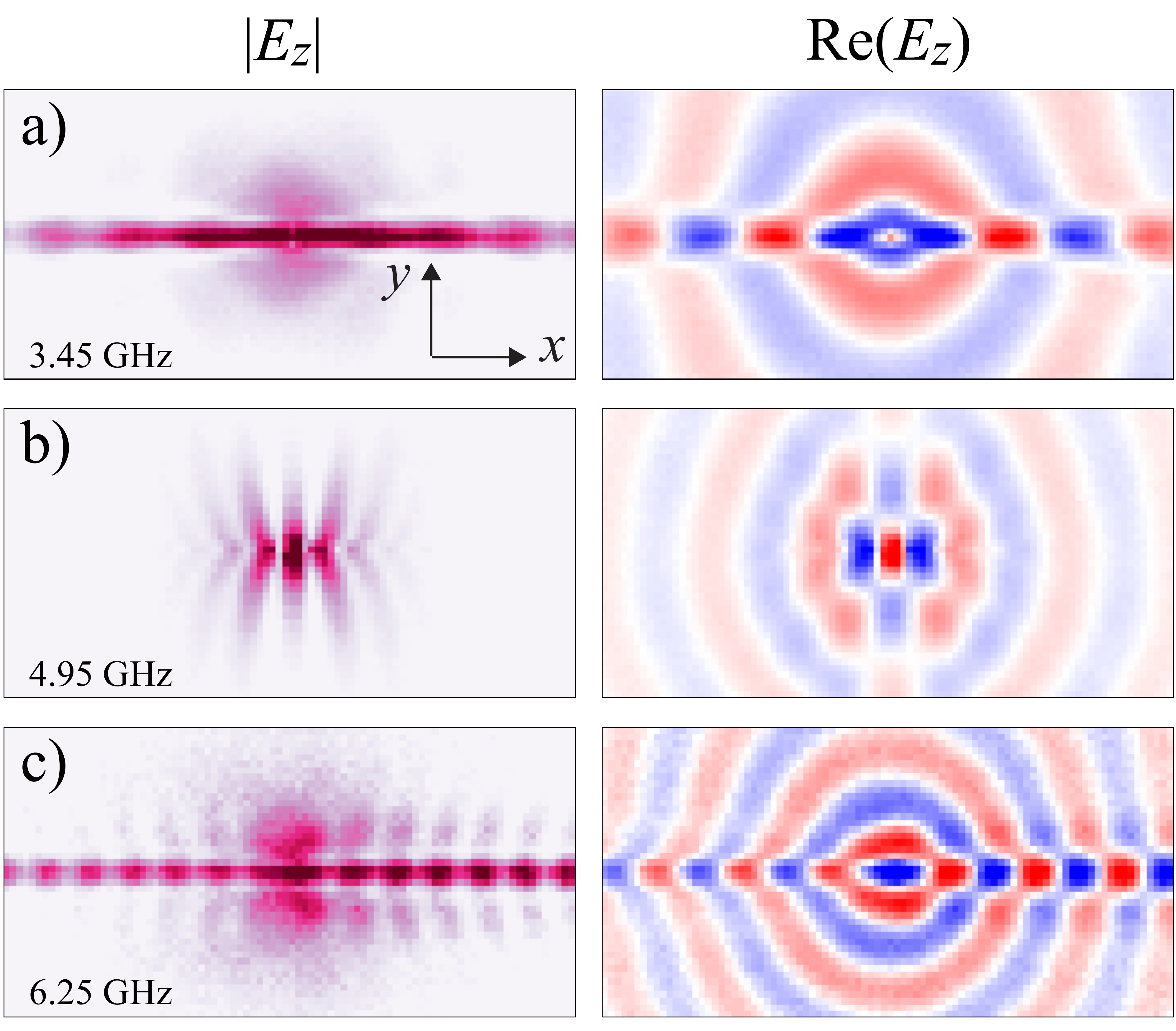}
  \caption{The spatial distribution of the absolute value (first column) and real part (second column) of the measured normal electric field component for the canalized quasi-TM surface waves. The scanning area of the structure consists of 48$\times$24 unit cells (336$\times$168~mm$^2$). }
  \label{fig6}
\end{figure}

The measured and simulated distributions  of the normal component of the magnetic field excited by a magnetic loop source (quasi-TE mode) and the reconstructed isofrequency contours are shown in Fig.~\ref{fig5}. The similar results for quasi-TM mode fields are shown in Fig.~\ref{fig6}.

\section{Results and discussion}

\subsection{Hyperbolic plasmons}

The intriguing consequence of Eq.~\eqref{Lorentz} and Eq.~\eqref{complementary_cond_Y} is that $\left[ \text{Im}(Y_x) \text{Im}(Y_y) \right] < 0$ for any frequency. Therefore, a self-complementary metasurface appears a genuine \textit{hyperbolic metasurface} in a resonant case (Fig.~\ref{fig2}a). Hence, the surface waves localized at a resonant self-complementary metasurface can be classified as the \textit{hyperbolic plasmon-polaritons}~\cite{yermakov2015hybrid,gomez2015hyperbolic,nemilentsau2016anisotropic}.

In a contrast to the conventional relatively narrow-band hyperbolic operational regime of two-dimensional systems caused by the resonances splitting, the broadband hyperbolicity of a self-complementary metasurface arising from Eq.~\eqref{complementary_cond_Y} is a unique feature and possesses a high-value practical potential. This result is in full agreement with the IFCs calculated analytically (Figs.~\ref{fig2}e,\ref{fig2}f,\ref{fig2}h), numerically and reconstructed experimentally (Fig.~\ref{fig5}i). Finally, the simulated (Fig.~\ref{fig5}c) and experimental (Fig.~\ref{fig5}m) field patterns are characterized by hyperbolic wavefronts.

Usually, hyperbolic metasurfaces strongly depend on the geometrical arrangement of the material constituents. The slight scaling of the unit cell sizes shifts the resonances and, as a consequence, the frequency range between the resonances corresponding to the hyperbolic regime. It means that even minor modifications, deformations and external effects can destroy the hyperbolic regime of a metasurface at a given frequency. This sharp dependence of the hyperbolic properties on the meta-atoms scaling and deformation can be denoted as \textit{extrinsic hyperbolicity}. The opposite situation takes place for the natural 2D materials such as black phosphorus, hexagonal boron nitride, van der Waals materials, because their hyperbolic properties do not depend on the structure size. In this sense, natural 2D materials possess \textit{intrinsic hyperbolicity}. The resonant self-complementary metasurface exhibits both kinds of hyperbolicity. First of all, it is still an artificial structure, so the scaling shifts the resonances. At the same time, the moderate resonances shifting for a self-complementary metasurface does not break the hyperbolic regime and, for instance, hyperbolic plasmons are still existing in contrast to the purely extrinsic hyperbolicity case. Therefore, we conclude that hyperbolicity of a resonant self-complementary metasurface can be classified as a quasi-intrinsic being an intermediate platform between conventional hyperbolic metasurfaces and natural hyperbolic materials.

Nevertheless, one should take into account that the hyperbolic-like dispersion of surface waves for real implementation can be significantly modified and even removed owing to the non-local contribution~\cite{gorlach2014effect,correas2015nonlocal,yermakov2018effective}.


\subsection{Switchable canalization direction}

We will denote canalization of surface waves as their collinear propagation and ultrafocused energy transport like those in Fig.~\ref{fig5} and Fig.~\ref{fig6}. The necessary conditions for the canalization are the strong anisotropy and near-zero admittance (conductivity) regime~\cite{yermakov2015hybrid,gomez2016flatland,correas2017plasmon}. The conditions $Y_y/Y_x \to \infty$ and $Y_x/Y_y \to \infty$ lead to the canalization along (Fig.~\ref{fig2}i) and orthogonal to (Figs.~\ref{fig2}g,~\ref{fig2}j) the strips, respectively. One can see that for a self-complementary metasurface this phenomenon is achieved in the vicinity of $Y_x$ and $Y_y$ resonances and can be found from the dispersion diagram as the surface plasmons resonances. However, from the experimental results in Fig.~\ref{fig6}, an effective canalization seems to occur only along the strips. The reason is related to the shifts of eigenvalues resonant frequencies caused by the non-perfect duality of the measured structure. Indeed, the condition of perfect zero-pole cancellation of the product between two admittance eigenvalues is violated for propagation orthogonally to the strips. In order to obtain a perfect canalization one should adjust the slots dimensions to compensate the dielectric substrate presence~\cite{baena2017broadband}. Besides, the vertical IFC is not purely flat due to the non-local interaction between the meta-atoms (Fig.~\ref{fig5}h). The analogous effect has been previously observed for waves propagating in a double mutually-orthogonal wire medium~\cite{simovski2004low}. In this case, the isofrequency contours are rather close to two orthogonal flat contours (4 straight lines), but transform to the 4 near-corner hyperbolic lines and a central closed contour due to the capacitive coupling between orthogonal wires. 

It is important to note that for the conventional anisotropic structure, the pole ($|\text{Im}(Y_y)| \gg 1$) or zero ($|\text{Im}(Y_y)| \ll 1$) of one eigenvalue (for instance, $Y_y$) is a necessary but not sufficient condition for the canalization. While for the self-complementary metasurface, the pole or zero behaviour of $Y_y$ automatically results in the opposite behaviour of $Y_x$ according to the Babinet's duality relation~\eqref{complementary_cond_Y}. For instance, $\text{Im}(Y_x)$ pole leads to $\text{Im}(Y_y) \approx 0$, so that one can observe the canalization along the strips ($|Y_x/Y_y| \to \infty$). 

The promising feature of the resonant self-complementary metasurface (even not perfectly self-complementary metasurface due to the presence of the dielectric substrate) is the possibility to switch between two orthogonal canalization directions by a small frequency shift. This effect is possible due to the existence of at least two consequent resonances of the metasurface. One of the eigenvalues has a zero at a characteristic frequency and a pole at a nearby frequency. The Babinet's principle guarantees that the reverse situation occurs for other eigenvalue. We demonstrate both numerically and experimentally the sharp switching between two orthogonal plasmon canalization directions in a narrow operational frequency range: (i) along the strips at 3.5~GHz (Fig.~\ref{fig5}l) and 3.45 (Fig.~\ref{fig6}a), (ii) across the strips at 5.1~GHz (Fig.~\ref{fig5}n) and 4.95 (Fig.~\ref{fig6}b), and again (iii) along the strips at 6.2~GHz (Fig.~\ref{fig5}o) and 6.25 (Fig.~\ref{fig6}c) for TE (Fig.~\ref{fig5}) and TM (Fig.~\ref{fig6}) polarization, respectively.

\subsection{Broadband polarization degeneracy}


According to Eq.~\eqref{kx_anal} the dispersions of two modes with orthogonal polarizations are identical for the surface waves propagation along the main axes directions. This feature is inherent to the bulk waves in any isotropic medium where the spectrum is doubly degenerate with respect to polarization. However, in general for surface waves the polarization degeneracy phenomenon of eigenmodes is achieved accidentally only at discrete frequency points. In our simple analytical model the broadband polarization degeneracy of the surface waves is fulfilled for the principal directions for any frequency (Fig.~\ref{fig2}b-\ref{fig2}c). Indeed, this result is in a good agreement with Ref.~\cite{gonzalez2015surface}, where it is shown that the dispersion curves of TM and TE surface waves localized at complementary L-type and C-type metasurfaces, respectively, are identical. 

\begin{figure}[htbp]
  \centering
  \includegraphics[width=0.97\linewidth]{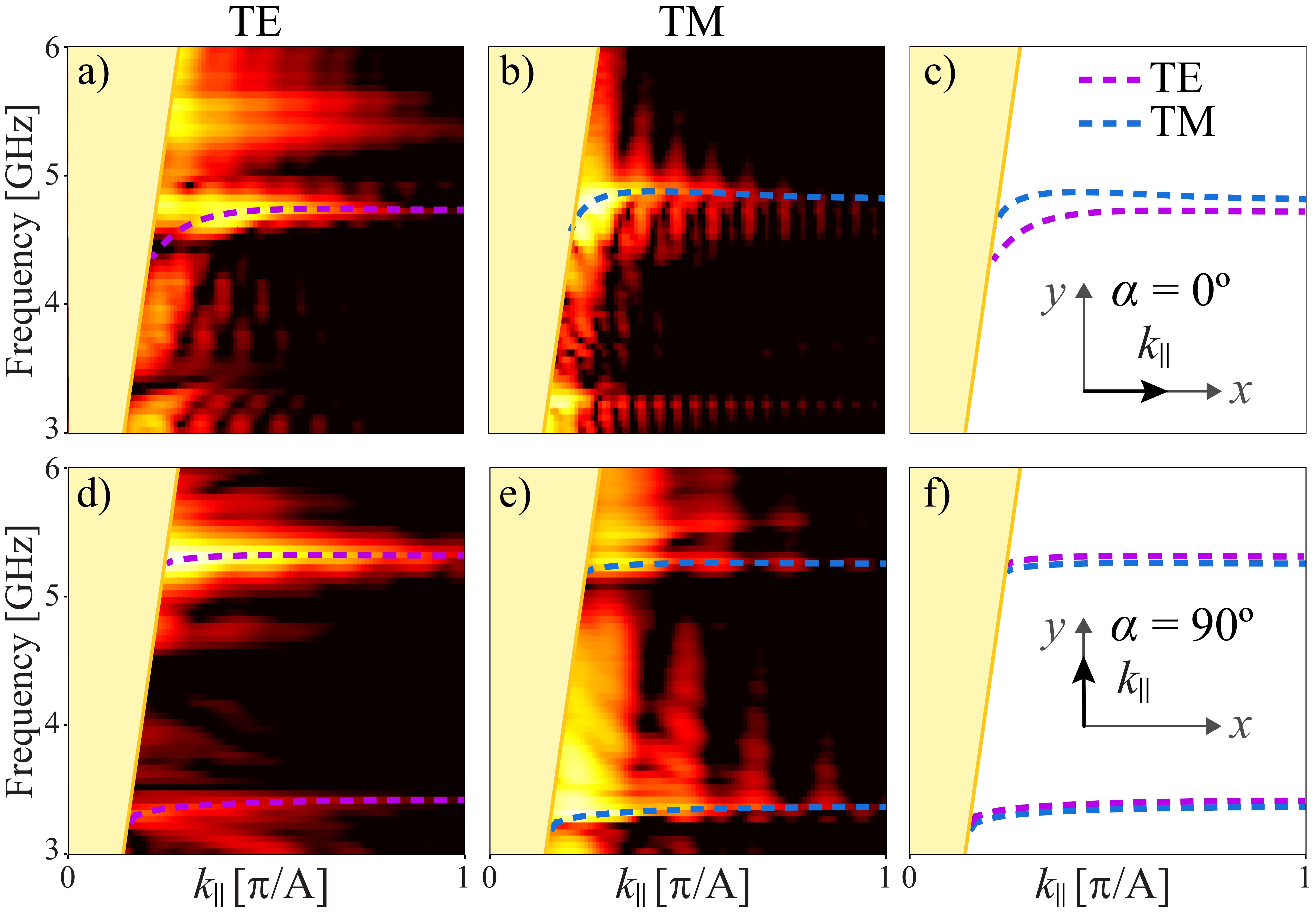}
  \caption{The reconstructed dispersion of surface waves excited by magnetic (a,d) and electric (b,e) sources propagating along (a,b) and across (d,e) the strips. The comparison of the reconstructed dispersion for TE and TM surface plasmons along (c) and across (f) the strips.}
  \label{fig7}
\end{figure}

Furthermore, the numerical results, shown in Figs.~\ref{fig4}h-\ref{fig4}k, confirm the polarization degeneracy for the sample, shown in Fig.~\ref{figUC}. Besides, we demonstrate the similar dispersion of TE and TM surface modes along and across the strips extracted from the measurements by using the recovered IFCs (Fig.~\ref{fig7}).

The demonstrated effect of broadband polarization degeneracy of surface waves was especially pronounced in the experiment for the canalization regimes. The frequencies corresponding to the canalization regime for quasi-TE (Figs.~\ref{fig5}l,~\ref{fig5}n,~\ref{fig5}o) and quasi-TM (Fig.~\ref{fig6}) modes are almost the same. Therefore, we state that our resonant  self-complementary metasurface in the vicinity of its resonances can support highly directional propagation of two orthogonally polarized surface waves with nearly the same phase velocity. It means that the canalized plasmon can be transmitted in the direction parallel to one of the main axes keeping the initial polarization of its source (e.g. circular, linear or elliptical polarization). Moreover, the effect of canalization allows for parallel routing of surface waves excited by multiple sources located nearby the same metasurface. This operational principle can be useful for photonic devices, while in the microwave range it can be directly applied to engineer the planar holographic leaky-wave antennas based on the modulated surface impedance with the capability to operate with arbitrary polarization in a multi-channel mode.


\section{Conclusions}

To conclude, we have studied both theoretically and experimentally the surface waves localized at a resonant self-complementary metasurface at microwaves. We have developed a consistent analytical model based on the effective surface admittance tensor approach. We have theoretically, numerically and experimentally demonstrated the wideband intrinsic hyperbolicity, canalization in frequency-switchable direction and broadband TE-TM polarization degeneracy of surface waves localized at a self-complementary metasurface. Importantly, we have shown the simultaneous canalization and TE-TM degeneracy of surface waves in the vicinity of the self-complementary metasurface resonances. It means that emitter energy can be transferred by virtue of surface waves highly directive and keeping the same polarization. The results obtained reveal that the self-complementary metasurfaces provide the platform for a number of applications including the polarization control and routing over optical and radiofrequency signal, on-chip devices, planar networks, photonic components, antennas and sensors.

\section*{Acknowledgments}

This work was supported by Russian Foundation for Basic Research (No. 20-02-00636). S.G. was supported by the President of the Russian Federation under grant MK-3620.2019.8. O.Y. acknowledges the support from Foundation for the Advancement of Theoretical Physics and Mathematics "BASIS".

\bibliography{apssamp}

\end{document}